\documentclass[showpacs,twocolumn,pre]{revtex4}
\usepackage{psfrag,epsfig,amsfonts,amssymb,amsmath,graphicx,slashbox}
\usepackage{dcolumn}

\newcommand{\boldmu}{\mbox{\boldmath $\mu$}}
\newcommand{\bolddelta}{\mbox{\boldmath $\Delta$}}

\begin{document}

\title{Optimal evaluation of single-molecule force spectroscopy experiments}
\author{Sebastian Getfert}
\author{Peter Reimann}
\affiliation{Fakult\"at f\"ur Physik, Universit\"at Bielefeld, 33615 Bielefeld, Germany}

\begin{abstract}
The forced rupture of single chemical bonds under external load 
is addressed. A general framework is put forward to optimally 
utilize the experimentally observed rupture force data for 
estimating the parameters of a theoretical model.
As an application we explore to what extent a distinction
between several recently proposed models is feasible
on the basis of realistic experimental data sets.
\end{abstract}
                                                 

\maketitle

{\em Introduction:}
Single-molecule force spectroscopy \cite{mer01}
refers to the experimental observation of chemical
dissociation by pulling apart the molecular complex 
of interest at a constant velocity $v$ 
until the bond breaks.
The evaluation of the resulting rupture 
force data is a non trivial task
\cite{eva97,rie98,mer99,str00,hey00,ngy03,rai06}:
for one and the same $v$,
the rupture forces are found to be 
randomly distributed over a wide range, 
and for different $v$,
different such probability 
distributions are obtained.
While both on the experimental and modeling sides
a great amount of work has led to substantial
progress and sophistication, much less effort 
has been spent to improve the still rather basic
methods of connecting and comparing theory and 
experiment. This is the subject of our present work.

Our starting point is the probability density 
$p_1(f|\boldmu,v)$ that a dissociation 
event occurs at a pulling force $f$,
given the pulling velocity $v$ and 
{\em any theoretical model} with certain 
model parameters $\boldmu$.
Now, our main question is:
What is the optimal estimate of those
model parameters $\boldmu$ that can be 
extracted from any given set of $N$ rupture 
forces ${\bf f} = \{f_i\}_{i=1}^N$ and pulling
velocities ${\bf v}=\{v_i\}_{i=1}^N$ ?
Since the $f_i$ are statistically independent, 
the probability of observing the given set of 
rupture forces $\bf f$ reads
\begin{equation}
p({\bf f} | \boldmu, {\bf v}) = 
\prod_{i=1}^N p_1(f_i|\boldmu,v_i) \ .
\label{1}
\end{equation}
The {\em main result} of our paper is that 
the optimal parameter estimate 
is obtained by simply {\em maximizing (\ref{1}) 
with respect to} $\boldmu$: 
no other ``recipe'' is able to yield estimates 
closer to the true parameter values systematically,
i.e., on the average over many data sets ${\bf f}$.

{\em Example:}
The explicit form of $p_1$ in (\ref{1}) depends on 
the specific model one is considering, and
similarly for the meaning and even the 
number of the model parameters $\boldmu$.
While our general theory applies to any
model, an illustrative and particularly 
simple example is provided by the most 
widely used model \cite{eva97,mer01,bel78}, 
viewing the dissociation as a rate process of 
the form $\dot n(t) = -k\left(f(t)\right)n(t)$,
where $n(t)$ denotes the bond survival probability 
and $k(f(t))$ the dissociation rate at 
the instantaneous pulling force $f(t)$, 
and adopting the approximations
\cite{eva97,mer01,bel78}
\begin{equation}
f(t) = \kappa v t\ , \ \ k(f)=\exp\{\lambda+\alpha f\} \ .
\label{2}
\end{equation}
Here, $\kappa$ is the net elasticity of the setup 
and $\kappa v$ is the so-called loading rate.
While they are considered as known \cite{eva97,mer01,bel78},
$\boldmu=(\lambda,\alpha)$ are the two unknown model parameters 
in the case of our specific example at hand.
Their physical meaning is discussed in detail, e.g., 
in \cite{mer01,eva97}:
$k(0)=\exp\{\lambda\}$ represents the force-free dissociation 
rate and $\alpha k_BT$ the dissociation length 
(distance between potential well and barrier along 
the reaction pathway), where $k_BT$ is the
thermal energy.
Having thus completely specified the model 
\cite{eva97,mer01,bel78}, a straightforward calculation 
yields for this particular example the explicit result
\begin{equation}
p_1(f|\boldmu,v)=
\frac{e^{\lambda+\alpha f}}{\kappa v}
\exp\left\{-\frac{e^{\lambda}}{\kappa v}\frac{e^{\alpha f}-1}{\alpha}\right\} \ .
\label{3} 
\end{equation}

{\em Formal analysis:}
The quantity in (\ref{1})
is called the {\em likelihood} and plays a 
central role in Bayes' theorem \cite{siv96}
\begin{eqnarray}
  p(\boldmu | {\bf f},{\bf v}) 
  =p({\bf f} | \boldmu,{\bf v})\, p(\boldmu, {\bf v}) / p({\bf f}, {\bf v}) \ .
  \label{4}
\end{eqnarray}
The left-hand side represents the ``likeliness'' of 
$\boldmu$, given the data ${\bf f}$,  ${\bf v}$, 
and hence is clearly of central interest for our purposes.
Considering also the right-hand side as a function of
$\boldmu$, it is equal to the likelihood from (\ref{1})
times $p(\boldmu, {\bf v})$, encapsulating all 
our knowledge about $\boldmu$
before the measurement, 
times a $\boldmu$-independent factor $1/p({\bf f}, {\bf v})$.
We emphasize that we will not use the Bayesian 
formalism in our actual calculations below, only
in their intuitive interpretation.

Next we exploit the fact that typically a
quite large set of rupture data ${\bf f}$ 
is available. Thus, focusing on large $N$, 
it is convenient to rewrite 
(\ref{1}) as
\begin{eqnarray}
p({\bf f} | \boldmu, {\bf v}) & = & \exp\{-N\,s_N({\bf f}, \boldmu , {\bf v})\}
\label{5}
\\
s_N({\bf f}, \boldmu , {\bf v}) & := & - N^{-1}\sum_{i=1}^N 
\ln p_1(f_i|\boldmu, v_i) \ .
\label{6}
\end{eqnarray}
Furthermore, we assume that the relative frequency
with which the different pulling velocities $v$ are sampled 
converges toward a well-defined limit 
$\rho(v)$ for $N\to\infty$.
Finally, we assume that the rupture forces $f_i$ have been 
sampled according to the ``true'' distribution $p_1(f_i|\boldmu_0, v_i)$
with unknown, {\em ``true'' model parameters} $\boldmu_0$.
Then it follows from the law of large numbers \cite{cov91} that
\begin{eqnarray}
s_N({\bf f}, \boldmu , {\bf v}) & \to & s(\boldmu):=
- \langle \ln p_1(f|\boldmu,v)\rangle
\label{7}
\end{eqnarray}
for $N\to\infty$, where $\langle\cdots\rangle$ indicates an 
average over $f$ and $v$ with weight $p_1(f|\boldmu_0,v)\, \rho(v)$. 
Hence, $s_N$ is an intensive, entropy like quantity. 
Observing that $s(\boldmu)-s(\boldmu_0)$ is a relative entropy 
of the form $\langle \ln [p_1(f|\boldmu_0,v)/p_1(f|\boldmu,v)]\rangle$,
one can infer \cite{cov91} that $s(\boldmu)$ has a unique 
absolute minimum at $\boldmu=\boldmu_0$. 
For any given ${\bf f}$ and ${\bf v}$, we denote by 
$\boldmu^\ast=\boldmu^\ast({\bf f},{\bf v})$ 
{\em the maximum of the likelihood} $p({\bf f} | \boldmu, {\bf v})$
with respect to $\boldmu$, or, equivalently, the minimum of 
$s_N({\bf f}, \boldmu , {\bf v})$ in (\ref{5}).
Since $s_N$ converges for large $N$ toward $s$ according to (\ref{7}),
also the minimum $\boldmu^\ast$ of the former converges 
to the minimum $\boldmu_0$ of the latter.
Consequently, for $\boldmu$ close to $\boldmu^\ast$
and large $N$, we can expand $s_N({\bf f}, \boldmu , {\bf v})$ 
up to second order about its minimum at $\boldmu^\ast$ and
the Hessian matrix of $s_N({\bf f},\boldmu^\ast,{\bf v})$ 
can be replaced by the Hessian $H=H(\boldmu_0)$ of $s(\boldmu_0)$, 
i.e.,
\begin{eqnarray}
s_N({\bf f},\boldmu^\ast+\bolddelta,{\bf v}) 
& = & s_N({\bf f},\boldmu^\ast,{\bf v}) +
\bolddelta^\dagger H \bolddelta/2 \ .
\label{8}
\end{eqnarray}
For large $N$ this is a very good approximation for all $\boldmu$-values 
with an appreciable weight in (\ref{5}), i.e.,  
\begin{equation}
p({\bf f} | \boldmu, {\bf v}) 
\propto
\exp\{-N(\boldmu-\boldmu^\ast)^\dagger H (\boldmu-\boldmu^\ast)/2\} \ .
\label{9}
\end{equation}
Within this narrow peak region,
the factor $p(\boldmu,{\bf v})$ in (\ref{4}), 
though usually unknown in detail,
can be considered as approximately 
constant, i.e., $p({\boldmu} | {\bf f}, {\bf v})
\propto p({\bf f} | \boldmu, {\bf v})$.
Given ${\bf f}$ and ${\bf v}$, the likelihood (\ref{1})
thus quantifies the ``likeliness'' that the 
``true'' model parameters are $\boldmu$.

Upon repeating the entire set of $N$ pulling 
experiments with the same set of pulling velocities 
${\bf v}$, a different set of rupture data ${\bf f}$ 
will be sampled, yielding a different maximum 
likelihood estimate $\boldmu^\ast$.
While the probability distribution of ${\bf f}$
is given by (\ref{1}) with $\boldmu=\boldmu_0$,
what can we say about the distribution 
of the maximum likelihood estimates $\boldmu^\ast$?
To determine its first moments, 
we differentiate (\ref{8}) and choose 
$\bolddelta=\boldmu_0-\boldmu^\ast$, resulting in
\begin{equation}
\boldmu^\ast-\boldmu_0 = - H^{-1}\partial s_N({\bf f},\boldmu_0,{\bf v})/\partial\boldmu \ .
\label{10}
\end{equation}
Averaging over ${\bf f}$ yields zero on the right-hand side, as can be
inferred from (\ref{6}), (\ref{7}) and the fact that $\boldmu_0$ is 
the minimum of $s$. Hence, 
\begin{equation}
\langle\boldmu^\ast\rangle = \boldmu_0 \ ,
\label{11}
\end{equation}
i.e., the maximum likelihood estimate is ``unbiased''.
An analogous but somewhat more involved calculation 
\cite{unp} yields for the second moments the result
\begin{equation}
\langle [\boldmu^\ast-\boldmu_0]\, 
[\boldmu^\ast-\boldmu_0]^\dagger\rangle = (N\, H)^{-1} \ .
\label{12}
\end{equation}
Observing that $(N\,H)^{-1}$ is the covariance 
matrix of the distribution from (\ref{9}),
we arrive at our 

{\em First main conclusion}:
For any given, sufficiently large data 
set ${\bf f}$, the expected deviation of 
the concomitant maximum likelihood
estimate $\boldmu^\ast$ from the ``true'' 
parameters $\boldmu_0$ immediately follows 
from the ``peak width'' of likelihood (\ref{1}),
considered as a function of $\boldmu$.

Similarly, from the higher moments one can infer \cite{unp} 
that $\boldmu^\ast$ is Gaussian distributed, yielding with
(\ref{9}) our

{\em Second main conclusion}:
Apart from the peak position and a normalization factor,
the likelihood (\ref{1}) for one given data set ${\bf f}$ 
looks practically the same as the distribution of the 
maximum likelihood estimates $\boldmu^\ast$ from many 
repetitions of the $N$ pulling experiments.

\begin{figure}
\epsfxsize 0.85 \columnwidth
\epsfbox{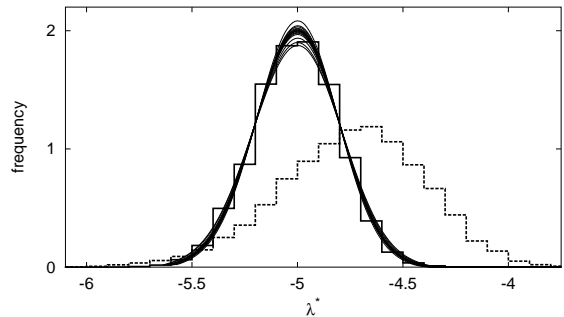}
\caption{
Solid histogram: Numerically determined distribution of the first 
components of the maxima $\boldmu^\ast=(\lambda^\ast,\alpha^\ast)$ 
of the likelihood (\ref{1}) for 10000 
``computer experiments''.
For each of them, $N=400$ rupture forces $f$ were 
sampled according to (\ref{3}), 100 for each of 
the four loading rates $\kappa v=50$,
$200$, $1000$, $5000$ pN/s
and with ``true''  parameters $\lambda_0=-5$ 
and $\alpha_0=0.1$ pN$^{-1}$.
These are typical numbers in 
``real experiments'' \cite{mer01}.
Thin lines: Likelihood (\ref{1}) for the first 
15 of the 10000 experiments after integrating 
over $\alpha$, shifting the maximum to $\lambda_0$,
and normalizing (some are almost indistinguishable).
Dotted histogram: Distribution of the estimates for $\lambda$
according to the ``standard method'', as described
in the main text.
}
\label{fig1}
\end{figure}

Figure 1 illustrates these findings by means of the 
example from (\ref{2}), (\ref{3}).
Since two-dimensional distributions
are difficult to compare graphically,
we focus on the marginal distributions 
for the first component $\lambda$ 
of $\boldmu=(\lambda,\alpha)$
(the findings for $\alpha$ are similar).
The close agreement of the
15 thin lines with the histogram
in Fig. 1 very convincingly 
illustrates our two conclusions above.

In view of the argument below (\ref{9}),
it seems intuitively quite plausible that 
the maximum of the likelihood $\boldmu^\ast$ 
should be the best possible guess for the 
unknown true parameters $\boldmu_0$.
A more rigorous line of reasoning
starts with an arbitrary ``recipe''
of estimating the true parameters $\boldmu_0$
from a given data set ${\bf f}$, formally
represented by some function
$\tilde\boldmu ({\bf f})$.
The only assumption is that this
recipe is unbiased, i.e., upon repeating the
same experiment many times, on the average, 
the ``true'' parameters are recovered, 
$\langle \tilde\boldmu ({\bf f})\rangle =\boldmu_0$.
By generalizing the well-kown Cram\'er-Rao 
inequality \cite{cov91}, which in turn is basically a 
descendant of the Cauchy-Schwarz inequality,
one can show \cite{unp} for any such ``recipe'' 
$\tilde\boldmu ({\bf f})$ that 
\begin{equation}
\langle [\tilde\boldmu-\boldmu_0]\, 
[\tilde \boldmu-\boldmu_0]^\dagger\rangle - (N\, H)^{-1} \geq 0
\ ,
\label{14}
\end{equation}
i.e., the matrix on the left-hand side is 
non-negative definite.
Comparison with (\ref{12}) yields our

{\em Third main conclusion:} There
is no unbiased estimator $\tilde\boldmu$
of the true parameters $\boldmu_0$ 
which on the average outperforms
the maximum likelihood estimate 
$\boldmu^\ast$.

\begin{figure*}
\epsfxsize 1.4 \columnwidth
\epsfbox{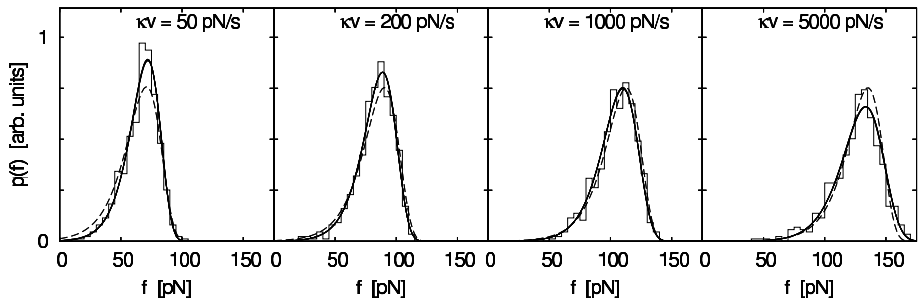}
\caption{
Rupture force distribution for different 
loading rates $\kappa v$. 
Histograms: numerically generated rupture forces 
according to (\ref{18}) with $\gamma=2/3$,
$\lambda_0=-5$, $\alpha_0=0.1$ pN$^{-1}$, $\epsilon_0=15$.
For each $\kappa v$, we sampled 500 forces, i.e., $N=2000$.
Solid: maximum likelihood fit $p_1(f|\boldmu^\ast,v)$
according to (\ref{18}) for $\gamma=1/2$ and
$\gamma=2/3$ (not distinguishable in this plot).
Dashed: same for $\gamma=1$.
Upon repeating the entire ``numerical experiment'',
the resulting plots always look practically 
the same.
}
\label{fig2}
\end{figure*}

The remaining possibility that a biased 
estimator may be even better is rather subtle 
to treat rigorously, but intuitively this 
seems quite unlikely.
Furthermore, in the above conclusion 
we exploited the relation (\ref{12}) which 
is strictly correct only for asymptotically 
large $N$.
Finally, the criterion of minimizing the
left-hand side in (\ref{14}) itself is
in principle also debatable, but hardly in 
practice.
Being unable to make any further progress 
along these lines, we directly compared the
maximum likelihood estimate with other known
``recipes'' of evaluating single-molecule rupture data 
\cite{eva97,mer01,mer99,str00,rai06,ngy03,evs03}.
In all cases we found that the maximum likelihood
was superior.

{\em Three case studies:}
1. In single molecule force spectroscopy,
the most widely used ``recipe''
for estimating parameters consist of the following steps:
(i) Fit a Gaussian to the observed rupture force
distribution for a fixed pulling velocity $v$
and approximate the most probable rupture force
$f^\ast$ by the maximum of that Gaussian.
(ii) Plot $f^\ast$ for different $v$
versus $\ln (v)$ and fit the resulting points 
by a straight line.
(iii) Assume that the model (\ref{2}), (\ref{3}) 
is applicable and deduce its model parameters
$\boldmu=(\lambda,\alpha)$ from the slope and the
axis intercept of the straight line as detailed,
e.g., in \cite{eva97,mer01,mer99,str00,evs03}.
We have applied this procedure to each
of the 10000 experiments in Fig. 1 and plotted
the distribution of the resulting estimates for
$\lambda$ in Fig. 1.
The conclusion is that the maximum likelihood
estimate represents a substantial improvement
compared to the so far ``standard method''
of data evaluation in this field.

2. Generalizations of the rate (\ref{2}) 
of the form
\begin{equation}
k(f)=(1-\gamma\alpha f/\epsilon)^{1/\gamma-1}\, 
e^{\lambda+\epsilon [1-(1-\gamma\alpha f/\epsilon)^{1/\gamma}]}
\label{16}
\end{equation}
with three model parameters ${\boldmu}=(\lambda,\alpha,\epsilon)$
have recently attracted considerable interest \cite{gen}.
Here, $\lambda$ and $\alpha$ have the same physical meaning 
as in Eq. (\ref{2}), $\epsilon:=E_b(0)/k_BT$ stands for the 
force-free activation energy barrier in units of the thermal energy $k_BT$, 
while $\gamma\in\{1/2,2/3,1\}$
labels three different models:
For $\gamma=1$ the parameter $\epsilon$ drops 
out and one recovers (\ref{2}),
$\gamma=2/3$ reproduces the Kramers rate for a cubic reaction potential, 
and $\gamma=1/2$ corresponds to a parabolic potential well 
with a cusp barrier \cite{gen}.
The resulting rupture force distribution
\begin{equation}
p_1(f| \boldmu, v) =
\frac{k(f)}{\kappa v}
\exp\left(
-\frac{e^{\lambda}}{\kappa v}
\frac{e^{\epsilon [1-(1-\gamma\alpha f/\epsilon)^{1/\gamma}]}
-1}{\alpha}
\right) 
\label{18}
\end{equation}
with $k(f)$ from (\ref{16}) can be determined analogously 
to (\ref{3}). 
There is an ongoing debate in the literature about
which of the three models is most appropriate to 
evaluate experimental rupture data \cite{gen}.
Taking for granted that one of the three models 
approximates the ``truth'' satisfactorily, choosing
$\boldmu=\boldmu^\ast$ is -- according to our above
conclusions -- the closest one can get to the 
``full truth'' on the basis of one given data set 
${\bf f}$.
In case of disagreement about the ``true'' 
$\gamma$-value, a fully objective selection criterion
seems impossible to define in principle.
In practice, the usual criterion is the comparison
with the basic ``true'' quantity observed experimentally,
namely, the distribution of rupture forces.
In view of Fig. 2, we conclude that under typical
experimental conditions it is absolutely 
impossible to decide whether 
$\gamma=1/2$ or $\gamma=2/3$ is ``better'', and even $\gamma=1$
performs almost as well \cite{sil03}.

\begin{figure*}
\epsfxsize 1.6 \columnwidth
\epsfbox{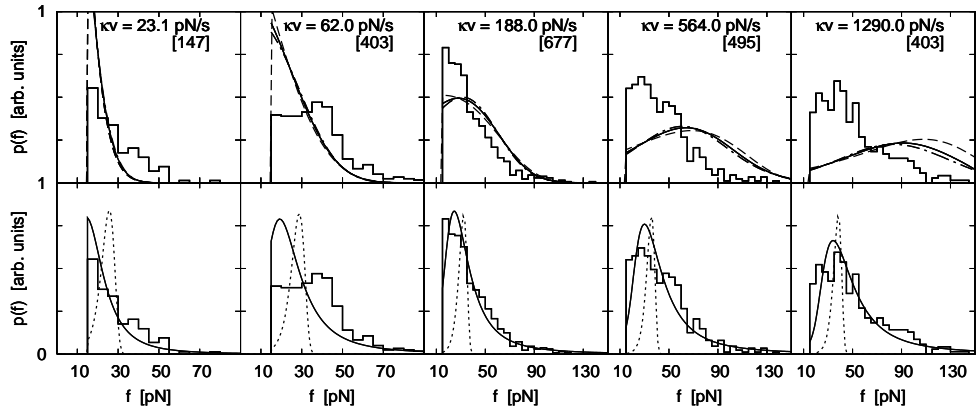}
\caption{
Same as Fig. 2 but for experimental rupture forces 
from \cite{ngy03}. The number of rupture events 
for each loading rate is indicated in brackets.
Upper row: Experimental data (histograms) and 
maximum likelihood fit $p_1(f|\boldmu^\ast,v)$ 
according to model (\ref{16}), (\ref{18})
for $\gamma=1$ (dashed), $\gamma=2/3$ (solid),
and $\gamma=1/2$ (dash-dotted).
Lower row: Experimental data (histograms), 
maximum likelihood fit $p_1(f|\boldmu^\ast,v)$ 
according to model (\ref{17}) (solid),
and best fit according to the ``standard method'', 
as described in the main text (dotted).
}
\label{fig3}
\end{figure*}

3. In Fig. 3 the same comparison as in Fig. 2 is
repeated, but now for real experimental data from 
\cite{ngy03}.
Again, the models (\ref{16}), (\ref{18}) 
with $\gamma=1/2$ and
$\gamma=2/3$ are hardly distinguishable;
$\gamma=1$ differs slightly more, while 
the ``standard  method'' yields a completely 
different ``best fit''.
However, none of them satisfactorily describes 
the ``experimental reality''.
The same incompatibility is recovered for all
other experimental data sets we analyzed so 
far, see also \cite{rai06}.
An almost perfect agreement (within the statistical
uncertainty of the experimental data) is obtained
by means of yet another recent extension \cite{rai06}
of (\ref{2}), (\ref{3}), 
considering the parameter $\alpha$ itself as 
randomly sampled from
\begin{equation}
\rho(\alpha)=
(2\pi\sigma_\alpha^2)^{-1/2}\,\exp\{-(\alpha-\bar\alpha)^2/2\sigma_\alpha^2\} \ ,
\label{17}
\end{equation}
resulting in a model with three parameters 
$\boldmu=(\lambda,\bar\alpha,\sigma_\alpha)$.
Possible reasons for such a heterogeneity of
the dissociation rate are uncontrollable variations 
of the experimental conditions or of the complicated 
biomolecular complex itself \cite{rai06}.

{\em Conclusions:} 
The maximum $\boldmu^\ast$ of the likelihood 
(\ref{1}) is the best possible estimate for the unknown 
model parameters $\boldmu$, given an appropriate model 
and a (sufficiently large) set of rupture forces ${\bf f}$.
The accuracy of this estimate follows from
the dispersion of the (approximately Gaussian)
likelihood peak about $\boldmu^\ast$.
The procedure is extremely simple and general.
For example, the pulling velocities $v_i$ may
be all the same, all different, or 
distributed in any other way, and the pulling 
force $f(t)$ may or may not increase linearly 
with time (only in the first case is there a
well defined loading rate $\kappa$; cf. (\ref{2})).

By means of a ``least-squares fit'' one gets -- by definition --
the best possible agreement between theory and experiment
with respect to any given ``deviation measure''.
A typical example is to optimize the agreement
between experimental and theoretical rupture 
force distributions.
In particular, the resulting agreement with the data in 
Figs. 2 and 3 would be (at least slightly) better than 
for any of the depicted theoretical lines.
However, our present goal is not to optimally fit rupture force 
distributions but rather to fit the unknown model parameters as 
closely as possible: they are the quantities of prime 
interest, and any other kind of fitting procedure is mainly
an intermediate step in order to estimate them. 
Using the rupture force distributions to fit parameters 
seems natural, but our paper shows that one can do better.
Our present comparison of the rupture forces in Figs. 2 and 3
serves a different purpose: 
once the parameters are estimated as well as possible according to our 
method, the resulting rupture force distributions can be used as an 
independent consistency test for a given hypothetical model.

\begin{center}
\vspace{-5mm}
---------------------------
\vspace{-4mm}
\end{center}
We thank R. Merkel for providing the experimental data
from \cite{ngy03}, M. Evstigneev and R. Ros for helpful
remarks, and the DFG for support under SFB 613.


\end{document}